\documentstyle[prl,aps,multicol]{revtex} 
\renewcommand{\narrowtext}{\begin{multicols}{2} \global\columnwidth20.5pc}
\renewcommand{\widetext}{\end{multicols} \global\columnwidth42.5pc}
\newcommand{\Lrule}{\vspace*{-0.2in}\noindent\vrule width3.5in height.2pt
  depth.2pt \vrule depth0em height1em}
\newcommand{\Rrule}{\vspace{-0.1in}\hfill\vrule depth1em height0pt \vrule
  width3.5in height.2pt depth.2pt\vspace*{-1pc}}

\input{epsf.tex}

\begin{document}

\bibliographystyle{prsty}

\draft 

\title{Distribution of the Absorption by Chaotic States in Quantum Dots}
\author{Nobuhiko Taniguchi$^{1}$ and Vladimir N. Prigodin$^{2}$}

\address{$^1$Department of Physical Electronics, Hiroshima University,
  Kagamiyama, Higashi-Hiroshima 739, Japan\\
  $^2$ Max-Planck-Institute f\"ur Physik komplexer Systeme, Au\ss enstelle
  Stuttgart, Heisenbergstr. 1, 70569 Stuttgart, Germany}

\date{\today}

\maketitle
\begin{abstract}
  The mesoscopic fluctuations of the absorption at optical transitions
  from a low energy regular state to high energy chaotic states in an
  aggregate of semiconductor quantum dots is studied.  We provide a
  universal dependence of the distribution of the absorption coefficient
  on the total number of dots and the ratio of the level broadening to the
  level spacing.  The distribution remain broad even at large broadening,
  and the absorption spectrum should demonstrate a strong sensitivity to
  weak magnetic field in the region of large and weak absorption.  The
  results can also apply to the absorption of Rydberg atoms in strong
  magnetic field at the pre-threshold ionization.
\end{abstract}

% Relevant PACS number:
%05.45.+b Theory and models of chaotic systems
%32.10.Dk Electric and magnetic moments, polarizability
%42.25.Bs Wave propagation; transmission and absorption (for radiowave
%         propagation, see 41.20.J and 84.40.C; for propagation in atmosphere, 
%         see 42.68.A; see also 52.40.D, N--in plasma physics)
%73.20.Dx Electron states in low-dimensional structures (superlattices,
%         quantum well structures and multilayers)
%73.23.-b Mesoscopic systems
%73.23.Ad Ballistic transport
%73.23.Hk Coulomb blockade; single-electron tunneling
%73.23.Ps Other electronic properties of mesoscopic systems
%78.20.Ci Optical constants: refractive index, complex dielectric constant,
%         absorption, reflection and transmission coefficients, emissivity
%78.66.-w Optical properties of specific thin films, surfaces, and
%         low-dimensional structures: superlattices, quantum well structures,
%         multilayers, and microparticles

\narrowtext

Semiconductor quantum dots are now a subject of intensive study as
potential devices with controlled optical
properties~\cite{AltshulerLeeWebb,Siegel93}.  Because of restricted
geometry, their spectra are discrete and are modified by variation of the
dot size.  The experimental absorption spectrum demonstrates a good
agreement with theoretical calculations for optical transitions between
the first few size-quantized levels~\cite{Ekimov88,Efros93}.  For these
low energy states, fluctuations of the dot size give broadening to
observed transitions, but their levels remain still well identified.
Fluctuations of the confinement potential play no prominent role since the
corresponding electron wave length is of the order of the dot size, i.e.,
exceeds a length scale of these irregularities.  For sufficiently high
energy states whose wave lengths amount to an atomic scale or a
characteristic scale of fluctuations, their levels and wavefunctions are
determined by such `imperfections' of the confinement potential.  The size
of the dot controls only the average interlevel spacing.  The confinement
part fluctuates from dot to dot and even for a given dot wavefunctions of
such high energy states changes unpredictedly from level to level.  This
allows us to take the statistical description for the high energy chaotic
states.

In the present paper, we study the absorption coefficient for the
transitions from a given initial state into these high energy chaotic
states.  The initial state can be the ground state or a low energy state
as well as a local electronic state, e.g., at the donor inside the dot.
For an aggregate of dots, the experimental absorption spectrum should show
featureless background with rare high peaks and the weak absorption
regions.  Our result describes the statistics of these fluctuations.  We
have found that the amplitude of fluctuations in the absorption depends
strongly on the total number of dots, and also on the ratio of the level
broadening to the level spacing.  When the latter is small, the
fluctuation of absorption spectra becomes highly asymmetric, demonstrating
abrupt decay for small ones and slow power-law decay for large
fluctuations. The fluctuations remain appreciable even when the broadening
is much larger than the level spacing and therefore the fluctuations
should be still observable for an ensemble of finite number of dots.

Similarly to the statistics of energy levels~\cite{MehtaHaake} and
wavefunctions in chaotic systems~\cite{PrigodinTaniguchi,Alhassid95}, the
distribution function for the absorption coefficient manifests the
universal behavior which is determined only by the generic symmetry of the
system~\cite{AltshulerSimons}.  In the absence of the spin-orbit
interaction, there are two universality classes present: the orthogonal
class for time-reversal-invariant (T-invariant) systems ($\beta=1$), and
the unitary class for time-reversal-breaking (T-breaking) systems
($\beta=2$).  The crossover between these two universality classes will
occur with the magnetic flux $\phi$ through a dot around $\phi\sim
h/e\sqrt{g}$ where $g$ is the dimensionless conductance~\cite{Dupuis91}.
We investigate the statistics of the absorption for both universality
classes.

Analogous absorption spectrum should be observed for Rydberg atoms in a
high magnetic field where the pre-ionization states are considered chaotic
\cite{Friedrich89}.

Consider the absorption for a single dot at optical transition from a
given initial state $| {\Phi _0} \rangle $ with the energy $E_0$ to
excited states $|\psi_n\rangle $ with energy $E_n$, which we assume fully
chaotic.  Under the resonance, the real part of the polarizability
$\alpha$ is negligible, so $\alpha(\Omega)$ becomes
\begin{equation}
  \alpha(\Omega) \approx i\pi\sum_{n} |P_{n0}|^2\; \delta_{\Gamma}(
  \Omega -\Omega_n),
\label{def:alpha}
\end{equation}
where $\Omega_n = E_n - E_0$, and $P_{n0}=\langle \psi_n|\hat
P|\Phi_0\rangle$, the transition dipole matrix element.  The explicit form
of the transition dipole $\hat P$ is not needed to proceed the following
argument.  $\delta_{\Gamma}$ is a $\delta$-function with a finite
broadening $\Gamma$, defined by $2\pi \delta _\Gamma (E) \equiv
{\Gamma/(E^2 +\Gamma^2/4})$.  A basic assumption here is that the level
width is fixed among these high energy excited states. Since such
broadening is dominated mainly by electron escape into the host matrix and
an electron can leave a dot through its whole boundary, fluctuations of
level widths should be suppressed.

The absorption (extinction) coefficient $\kappa$ in this regime is
determined by
\begin{equation}
  \kappa(\Omega) = \sqrt{2\pi |\alpha(\Omega)|}.  
\label{def:kappa}
\end{equation}

Reflecting random nature of $\Omega_n$ and chaotic wavefunctions,
$\alpha(\Omega)$ and $\kappa(\Omega)$ become highly fluctuating
quantities, either from sample to sample or by slightly changing $\Omega$.  
To examine these statistical properties, we will investigate the following
the probability distribution functions:
\begin{mathletters}
\begin{eqnarray}
  &&{\cal P}(x;\{\alpha\}) = \left\langle \delta(x-\alpha(\Omega)/\alpha_0)
\right \rangle ,\label{def:P-alpha}\\ &&{\cal P}(x;\{\kappa\}) = \left\langle
\delta(x-\kappa(\Omega)/\kappa_0) \right \rangle ,
  \label{def:P-kappa} 
\end{eqnarray}
\end{mathletters}
where $\alpha_0$ and $\kappa_0$ are the average values of $\alpha$ and
$\kappa$, respectively (see Eq.(\ref{alpha0}) below).  We obtain the
analytical expressions of ${\cal P}(x;\{\kappa\})$ and ${\cal
  P}(x;\{\alpha\})$, both for T-invariant and T-breaking systems.

We proceed this task by connecting the distribution functions of $\alpha$
or $\kappa$ with that of the local density of states $\nu (\bbox{r})$
defined by
\begin{eqnarray}
  &&\nu (\bbox{r},E) = \sum_{n} |\psi_{n}(\bbox{r})|^2
  \delta_{\Gamma}(E-E_{n}),  \label{def:nu}\\
  &&{\cal P}(x;\{\nu\}) = \left\langle {\delta (x-\nu (\bbox{r},E)/ \nu_0 )}
\right\rangle ,
\label{def:P-nu}
\end{eqnarray}
where $\nu_0 = \langle \nu(\bbox{r},E) \rangle$.  By utilizing statistical
properties of chaotic wavefunctions, we can show that
\begin{mathletters}
\label{Palpha-Pkappa}
\begin{eqnarray}
  &&{\cal P}(x;\{\alpha\}) = {\cal P}(x;\{\nu\}) \equiv {\cal P}(x),
\label{Palpha-P}\\ 
&&{\cal P}(x;\{\kappa\}) = 2x {\cal P}(x^2).
\label{Pkappa-P}
\end{eqnarray}
\end{mathletters}\noindent

First we show how Eq.~(\ref{Palpha-Pkappa}) was derive.  The average of
the polarizability become independent of $\Omega$ and is given by
\begin{equation}
\alpha_0 = {i\pi \over \Delta}\: \left\langle\,{|P_{n0}|^2}\right\rangle
\label{alpha0}
\end{equation}
where $\Delta=1/\nu_{0} V$ is the mean level spacing with the dot volume
$V$. $\langle\,{|P_{n0}|^2}\rangle$ is the average intensity of the
transition dipole moment~\cite{Taniguchi95} defined by
\begin{equation}
\left\langle\,{|P_{n0}|^2}\right\rangle = {1 \over V}\int \!\! d\bbox{r}_1
d\bbox{r}_2 \,
\langle \Phi_0 
  |\hat P| \bbox{r}_1\rangle f(r_{12}) \langle \bbox{r}_2|\hat P|
  \Phi_0\rangle,
\end{equation}
with $f(r_{12}) \equiv V \left \langle
\psi_{n}(\bbox{r}_1)\psi_{n}^{*}(\bbox{r}_2)\right\rangle$ to describe
spatial correlations of the wavefunction amplitude.  To calculate the
distribution, we need its higher moments
\begin{eqnarray}
  \left\langle {\,|P_{n0}|^{2k}} \right\rangle &&=
\int \prod_{i=1}^k \left( d\bbox{r}_id\bbox{r}'_i\,
  \left\langle {\Phi _0} \right|\hat P\left| {\bbox{r}_i} \right\rangle 
\left\langle {\bbox{r}'_i} \right|\hat P\left| {\Phi _0} \right\rangle \right)
\nonumber \\
&&\quad \times\left\langle \psi_n (\bbox{r}_1) \psi_n^*(\bbox{r}'_1)\cdots
  \psi_n 
  (\bbox{r}_k) \psi_n^*(\bbox{r}'_k) \right \rangle .
\end{eqnarray}
The spatial correlation between the local values of wavefunction exists only
within the mean free path $\ell$.  Beyond such distance, the wavefunction can
be seen to be fluctuating independently.  As a result, in the leading order of
$\ell \lambda^{d-1}/V$ ($\lambda$ is the wave length), we get
\begin{equation}
  \left\langle\, |P_{n0}|^{2k} \right\rangle = \left\langle \,|P_{n0}|^2
  \right\rangle^k \; \left\{
    \begin{array}{cc}
   k! & (\mbox{for }\beta =2)\\ {(2k-1)!!}&{(\mbox{for }\beta =1)}
\end{array}\right. .
\end{equation}
This shows that, as well as $|\psi_n(\bbox{r})|^2$, the distribution of
$|P_{n0}|^2$ and equally oscillator strength $f_n =2m\Omega_n |P_no|^2$
are characterized by the the Porter-Thomas
distribution~\cite{PorterThomas} after an appropriate rescaling.  By
comparing Eqs.~(\ref{def:alpha}) and (\ref{def:nu}), we can conclude
Eq.~(\ref{Palpha-P}), then Eq.~(\ref{Pkappa-P}) follows from
Eq.~(\ref{def:kappa}).

To evaluate ${\cal P}(x)$ analytically, it is convenient to work on its
Laplace transformation ${\cal L}(s)$~\cite{EfetovBeenakker}
\begin{equation}
  {\cal L}(s) =\int_0^\infty {dx\,e^{-s\gamma x}\; {\cal
      P}(x)},  \label{Laplace}
\end{equation}
where $\gamma \equiv \pi \Gamma / \Delta $.
The averaging for ${\cal L}(s)$ can be decomposed into averaging over
eigenfunctions and energy levels, and the former is recasted by the
Porter-Thomas distribution.  Since the Porter-Thomas distribution depends
on the symmetry parameter $\beta$, the result of the integration over
eigenfunctions ${\cal L}(s) =
{\cal L}_{\beta}(s)$ becomes
\begin{equation}
  {\cal L}_{\beta}(s)\equiv \left\langle \left({ {\det
      [(E-H)^2+\Gamma^2/4]\over \det[(E-H)^2+\tilde \Gamma^2 /
      4]}}\right)^{\beta/2}\; \right\rangle ,
\label{def:L_beta}
\end{equation}
where $H$ is the Hamiltonian of the system and $\tilde \Gamma =\Gamma
\sqrt {1+4s/\beta}$.  (We introduce $\tilde \gamma \equiv \pi\tilde
\Gamma/\Delta$ for later use.)

Eq.~(\ref{def:L_beta}) can be evaluated by the supermatrix $Q$
method~\cite{Efetov83}. 
A trick is needed for $\beta=1$ to generate $\left[ {\det \left(
  \cdots \right)} \right]^{1/ 2}$ in the numerator from the Grassmann
integration.  This can be achieved by using the form
\[
  {\cal L}_{o}(s)=\left\langle {\det[(E-H)^2+\Gamma^2/4] \over {
      {\sqrt{\det [(E-H)^2+ \Gamma^2 /4][(E-H)^2+\tilde \Gamma^2/4]}}}}
\right\rangle .
\]
which can be readily expressed within the $Q$-matrix
formalism~\cite{Efetov83}. After completing the mapping, 
Eq.~(\ref{def:L_beta}) is found to be equal to
\begin{equation}
  {\cal L}_{\beta}(s) = \int\!\! dQ\: \exp \left\{ {\rm
      STr}\left[(a_0\Lambda+a_1 \Lambda_s +a_2 C_B) Q\right] \right\}
\label{LinQ-matrix}
\end{equation}
where $\Lambda = {\rm diag}({\bf 1}_4,-{\bf 1}_4)$, $\Lambda_s = {\rm
  diag} ({\bf 1}_2,-{\bf 1}_2,-{\bf 1}_2,{\bf 1}_2)$, and $C_B ={\rm diag}
({\bf 0}_2,\sigma_1, {\bf 0}_2, -\sigma_1)$ ($\sigma_i$ are $2\times 2$
Pauli matrices).  The definitions of ${\rm STr}$ and integration of $Q$ as
well as the explicit structure of
$Q$-matrix can be found in~\cite{Efetov83}.  
The coefficients $a_i$ ($i=1,2,3$) are for the unitary case,
\begin{equation}
  a_0 = (\gamma +\tilde \gamma )/ 8;\quad a_1 = (\gamma -\tilde
  \gamma )/ 8; \quad a_2=0,
\end{equation}
and for the orthogonal, 
\begin{equation}
  a_0=(3\gamma +\tilde \gamma)/16;\quad 2a_1=-a_2=(\gamma -\tilde \gamma)/8.
\end{equation}

Evaluating\/ ${\cal L}(s)$ and ${\cal P}(x)$ for the unitary universality
class ($\beta=2$) was already done in the framework of the local density
of states distribution~\cite{EfetovBeenakker}.  Their results read
\begin{eqnarray}
  {\cal L}_{u}(s) && ={{e^{-\tilde \gamma }} \over {4\tilde \gamma \gamma
      }}\left[ {(\gamma +\tilde \gamma )^2e^\gamma -(\gamma -\tilde \gamma
    )^2e^{-\gamma }} \right]\label{L_u}\\
{\cal P}_{u}(x)&& =\sqrt {{\gamma \over {8\pi x^3}}}\exp \left[ {-\gamma
  \left( {x+ 1/x} \right)}/2 \right] \nonumber \\ && \qquad \times\left[
  {2\cosh \gamma +\left( {x+1/x-1/\gamma } \right)\sinh \gamma } \right].
  \label{P_u}
\end{eqnarray}
By using Eqs.~(\ref{Palpha-Pkappa}), we have the distribution of the
absorption coefficient for the unitary case.

To evaluate ${\cal P}_{o}(x)$ or ${\cal L}_{o}(s)$ is much more laborious
but still durable.  Technically the difficulty results from expanding the
exponent of Eq.~(\ref{LinQ-matrix}) and taking the highest order Grassmann
term to complete the integration $dQ$.  After straightforward but lengthy
calculations, we found that ${\cal
  L}_{o}(s)$ is given by  
\begin{eqnarray}
  {\cal L}_o(s) = e^{(\gamma -\tilde \gamma )/ 2} &&+ 2\int e^{-{1\over
      2}(\gamma +\tilde \gamma)\lambda _1\lambda _2+\gamma \lambda}
  \nonumber\\ &&~\times {\left[ {F(\lambda _1,\lambda _2,\lambda )}
  \right]^2 {d\lambda_1 d\lambda_2 d\lambda} \over {(\lambda _1^2+\lambda
    _2^2+\lambda ^2-2\lambda \lambda _1\lambda _2-1)^2}},
\label{L_o}
\end{eqnarray}
where the integral region are defined by
$\lambda_1,\lambda_2 \in (1,\infty)$ and $\lambda \in (-1,1)$, and with
$\mu_i = \sqrt{\lambda_i^2-1}$ (for $i=1,2$),
\begin{eqnarray}
  &&F(\lambda_1,\lambda_2,\lambda ) = {\tilde \gamma-\gamma \over 4 \pi}
  \int^{1}_{-1}\!\! dp\,\varphi(p)\: e^{\mu_1 \mu_2 p(\tilde \gamma -
    \gamma)/4},
\label{def:F}\\
  &&\varphi(p) = (\lambda_1 \lambda_2+\mu_1 \mu_2 p -
\lambda)/\sqrt{1-p^2}.
\label{def:varphi}
\end{eqnarray}
${\cal P}_{o}(x)$ is obtained by completing the inverse Laplace
transformation of Eq.~(\ref{Laplace}).  Evaluating the integral by
deforming the integration contour leads to
\widetext
\Lrule
\begin{equation}
  {\cal P}_o(x)=\sqrt {{\gamma \over {4\pi x^3}}}\left\{ e^{-{\gamma\over
      4x}(x-1)^2} +\int\!\!  d\lambda\prod_{i=1}^{2}d\lambda_i dp_i \;e^{
    -\gamma (\lambda _1\lambda _2-\lambda )-\gamma (\zeta - x)^2/4x}\,
  K(\lambda _1,\lambda _2,\lambda ,p_{1,}p_2)\right\}
\label{P_o}
\end{equation}
where the integral kernel $K$ is 
\begin{eqnarray}
  &&K(\lambda _1,\lambda _2,\lambda ,p_{1,}p_2)={\gamma \over
    {8\pi ^2x^2}}{\varphi (p_1)\varphi (p_2)} {{\gamma \zeta^3-2\gamma
      x\zeta^2-6x\zeta+\gamma x^2\zeta+4x^2} \over {(\lambda _1^2+\lambda
      _2^2+\lambda ^2-2\lambda \lambda _1\lambda _2-1)^2}},
  \label{def:K}\\
  &&\zeta(\lambda_1,\lambda_2,\lambda,p_1,p_2) = \lambda _1\lambda _2+\mu
  _1\mu _2{(p_1+p_2)/2}. \label{def:zeta}
\end{eqnarray}
\Rrule
\narrowtext\noindent
Combining Eqs.~(\ref{P_u},\ref{P_o},\ref{def:K}) with
Eq.~(\ref{Palpha-Pkappa})  consist of our main results in the paper.

The results for T-invariant systems (the orthogonal class) are
particularly interesting since this is the usual symmetry in observing
absorptions in quantum dots.  We also remark that ${\cal P}(x)$ can be
observed as the local density of states distribution in nuclear magnetic
resonant (NMR) experiments.  NMR experiments are performed rather within
the T-invariant situation, though only the analytical expression was known
for the unitary case so far~\cite{EfetovBeenakker}.  We emphasize
that our obtained result Eq.~(\ref{P_o}) serves in this respect as well.

Next we examine various asymptotic behavior of ${\cal P}_{o}(x)$.  For
$\gamma \gg 1$, ${\cal P}_{o}(x)$ can distribute only around the unity,
otherwise it's suppressed exponentially.  The dominant behavior for $1/
\gamma \ll x\ll \gamma$ is characterized by
\begin{equation}
  {\cal P}_{o}(x)\cong \sqrt {{\gamma \over {4\pi x^3}}}{\exp
    \left[ {-{\gamma \over 4}(\sqrt x-{1 \over {\sqrt x}})^2} \right]}.
\end{equation}
The form obtained above describes multilevel absorption.  The number of
levels which contribute to the absorption is of the order $\gamma$.  The
individual contribution are random because of different wave function and
therefore the matrix elements are uncorrelated.  In accordance with the
central limit theorem, the distribution function is obtained to be
Gaussian with width $\propto 1/\sqrt{\gamma}$.

For $\gamma \ll 1$, we can decompose the behavior of ${\cal P}_o(x)$ into
the three region: (1)~$x\ll \gamma \ll 1/ \gamma$, (2)~$\gamma \ll x\ll 1/
\gamma $, and (3)~$\gamma \ll 1/ \gamma \ll x$.  The evaluation of the
asymptotic behavior from the analytical expression leads to
\begin{equation}
  {\cal P}_o(x) \cong \sqrt{\gamma \over 4 \pi x^3}\times \left\{
  \begin{array}{ll}
    C e^{-\gamma/4x} &\quad{(x\ll \gamma \ll 1/ \gamma )}\\ {2 / \pi }
    &\quad{(\gamma \ll x\ll 1/ \gamma )}\\ C e^{-\gamma x/4}
    &\quad{(\gamma \ll 1/ \gamma \ll x)}
  \end{array}
\right.
\label{small-g-behavior}
\end{equation}
where $C$ is a numerical constant.  
Let us note that the distribution function Eq.~(\ref{small-g-behavior}) is
very asymmetric.  The maximum ${\cal P}_o(x;\{\alpha\}) = {\cal
  P}_{o}(x;\{\nu\})$ lies at $x \approx \gamma \ll 1$ and on the right
side from the maximum the function decays by power law as $x^{-3/2}$ in
contrast with $x^{-5/2}$ in the unitary case.  The left tail of ${\cal
  P}_o(x;\{\nu\})$ decay exponentially as $\exp(-\gamma/4x)$.  The
absorption at $\gamma \ll 1$ is determined by the rare single level,
thereby the distribution occurs to be shifted to small values.  The power
decay is formed by the spectral fluctuation.  At the same time far right
and left tails is due to the fluctuation of the matrix element of
wavefunction.

In the final, consider the absorption from a system of $N$-uncoupled dots
which have almost identical volumes but different shapes.  We can write
down the distribution function of the absorption from such a system by
\[
  {\cal P}^{(N)} (x) = \left\langle \delta (x - \sum_{j=1}^N
  {x_j\over N})\right\rangle
= \gamma \int_{-i\infty}^{+i\infty}
  \!\!{ds \over 2\pi i } e^{s\gamma x}{\cal L}^{N}({s\over N}) .
\]
In the unitary case where ${\cal L}(s)$ is given by Eq.~(\ref{L_u}), we
can write down $P^{(N)}$ explicitly,
\begin{eqnarray}
  {\cal P}^{(N)}_{u}(x) &&= \sqrt{N \gamma \over 2 \pi x} e^{-{N
      \gamma\over 2} (x+1/x)} \sum^{N}_{k=0}\sum^{k}_{m} {N\choose
    k}{k\choose m}\nonumber \\ &&\qquad \times
  H_{2m-k+1}(\textstyle{\sqrt{N\gamma \over 2x}}) \:\displaystyle{
    {(\cosh\gamma)^{N-k} (\sinh\gamma)^{k} \over 2^k (2 N\gamma
      x)^{2m-k+1}}}
\end{eqnarray}
where $H_{k}$ is the Hermite polynomials and its analytical continuation
into negative $k$.  In Fig.~1, we present the distribution of the
absorption coefficient ${\cal P}^{(N)}_{u}(x;\{\kappa\}) = 2x {\cal
  P}^{(N)}_{u}(x^2)$ for $\gamma=0.1$ and $10$ by changing $N$. 
When $\gamma \gg 1$ (Fig.~1a), the distribution ${\cal P}^{(N)}_{u}(x)$ is
nearly Gaussian around $x=1$, but with slight asymmetry.  Even for
$\gamma=10$ and $N=50$, we see the fluctuations amounts to the order of 
10 \%.   
The behavior for $\gamma \ll 1$ is quite different (Fig.~1b).  For $\gamma
\ll 1$ and $N=1$, there is a peak around $x\sim \gamma$ with strong
asymmetry, and the peak position move gradually to $1$ by increasing $N$.

The $N$-dependence of ${\cal P}^{(N)}(x)$ in the orthogonal case is
qualitatively very similar to that for the unitary case, except for the
different power-law decay of the tails.

In conclusion, we have studied the statistical properties of the
absorption in an aggregation of $N$-uncoupled quantum dots at the
transition between a low energy regular state and high energy chaotic
states.  The transition matrix elements and the oscillator strengths are
shown to obey the Porter-Thomas distribution, thus the statistics of the
polarizability has turned out to be identical to the local density states
distribution.  We have found that statistics of large deviation of the
absorption from its average value are strongly distinguished between
systems with and without T-invariance.  Therefore application of a weak
magnetic fields should have a pronounced effect on
the absorption spectrum by suppressing large fluctuations.

%% Acknowledgements
The authors are grateful to B.~L.~Altshuler, Al.~L.~Efros and T.~Ishihara for
fruitful discussions and their interest.  They are also very thankful the NEC
Research Institute for hospitality where this work was started.

%
% References 
% 
% To use BibTeX, 
%\bibliography{ref}

% Figures
% The next line should be commented out unless you have psfile(s).

\begin{figure}
%\centerline{\epsfxsize=7.5cm \epsfbox{ndots1.eps}}
%\centerline{\epsfxsize=7.5cm \epsfbox{ndots2.eps}}
\centerline{\epsfxsize=0.9\columnwidth \epsfbox{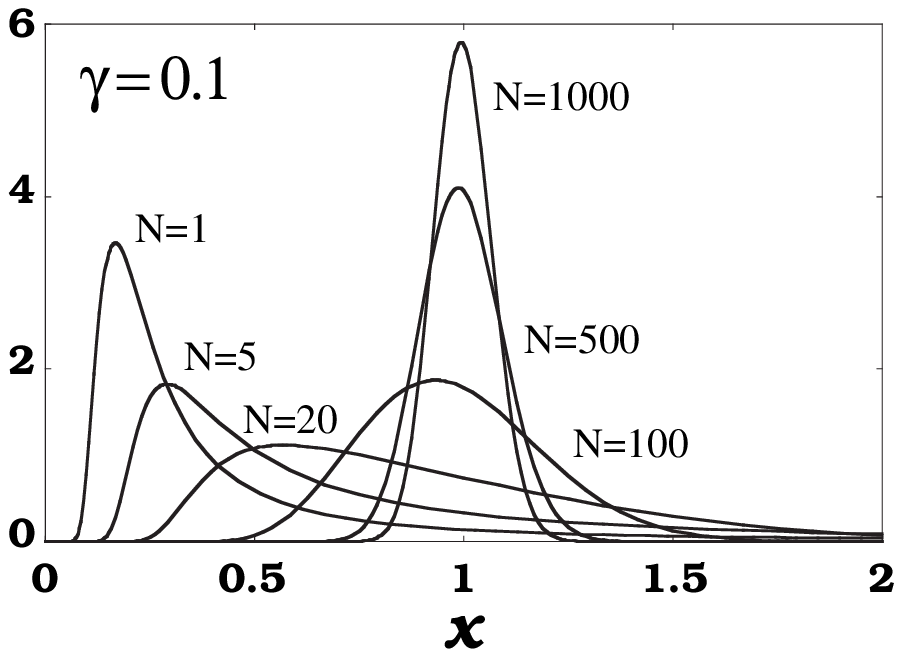}}
\centerline{\epsfxsize=0.9\columnwidth \epsfbox{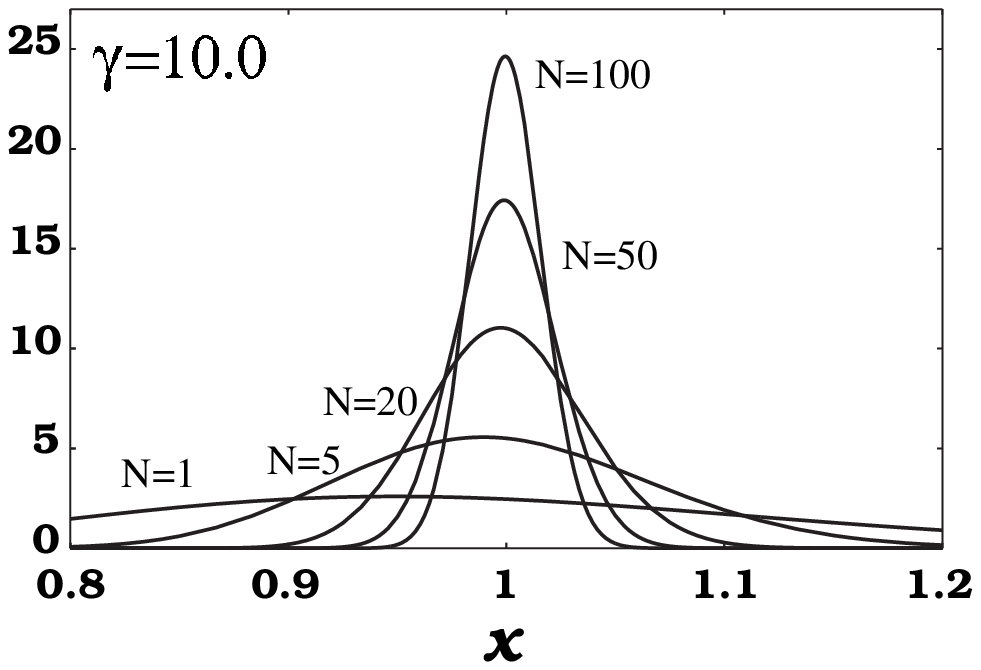}}
\caption{Distribution functions ${\cal P}^{(N)}_{u} (x;\{\kappa\})$  of the
  absorption coefficient in the unit of its average value for
  $N$-uncoupled dots.  Plottings are shown for (a) $\gamma=0.1$ and (b)
  $\gamma=10$, where $\gamma = \pi\Gamma/\Delta$ and $\Gamma$ and $\Delta$
  are the level width and the mean level spacing. }
\label{fig:N-dots}
\end{figure}

\widetext
\end{document}